\documentclass[12pt]{article}
\usepackage{amsmath}
\usepackage{graphicx,psfrag,epsf}
\usepackage{enumerate}
\usepackage{natbib}

\newcommand{\blind}{0}

\usepackage{color}
\newcommand{\etal}{\it et al.}

\begin{document}

\bibliographystyle{natbib}

\def\spacingset#1{\renewcommand{\baselinestretch}%
{#1}\small\normalsize} \spacingset{1}


\if0\blind
{
  \title{\bf Automatic Methods for Handling Nearly Singular Covariance Structures Using the Cholesky Decomposition of an Indefinite Matrix}
  \author{
  John R. Smith and Milan Nikolic\\
Physics Department\\ 
University of California Davis\\
Davis, California 95616\\ 
 and \\
Stephen P. Smith, Adjunct Faculty\\
Division of Mathematics and Science\\
Holy Names University\\
Oakland, California, 94619}
  \maketitle
} \fi

\if1\blind
{
  \bigskip
  \bigskip
  \bigskip
  \begin{center}
    {\LARGE\bf Title}
\end{center}
  \medskip
} \fi

\bigskip
\begin{abstract}
Linear models have found widespread use in statistical investigations. 
For every linear model there exists a matrix representation for which the ReML (Restricted Maximum Likelihood) 
can be constructed from the elements of the corresponding matrix. 
This method works in the standard manner when the covariance structure is non-singular. 
It can also be used in the case where the covariance structure is singular, 
because the method identifies particular non-stochastic linear combinations of the observations which must be constrained to zero. 
In order to use this method, the Cholesky decomposition has to be generalized to symmetric and {\it indefinite} matrices using complex arithmetic methods.
This method is applied to the problem of determining the spatial size (vertex) for the Higgs Boson decay in the ${\rm Higgs}\rightarrow 4~{\rm lepton}$ channel. A comparison based on the $\chi^2$ variable from the 
vertex fit for Higgs signal and $t\overline{t}$ background is presented and shows that the background can be greatly suppressed using the $\chi^2$ variable. One of the major advantages of this novel method over the currently adopted technique of b-tagging \citep{Tomalin 2008}\ is that it is not affected by multiple interactions (pile up). 
\end{abstract}

\noindent%
{\it Keywords:} Higgs Boson, Vertexing, b-tagging, ReML, Cholesky Algorithm, pile up, Singular Error Matrices\\

\spacingset{1.45}




 
\section{Introduction}
\label{sec:intro}
This paper describes an original automatic method for constructing constraints necessary to establish a consistent maximum likelihood problem in the presence of a nearly singular covariance structure based on the Cholesky decomposition of an {\em indefinite} matrix. The method is applied to the problem of constructing a vertex for extrapolated tracks in the context of a high energy physics data analysis. Track extrapolations involve a nearly singular covariance structure due to observational bias which affects high transverse momentum tracks. This observational bias is inherent in all track fitting algorithms used to reconstruct charged particle trajectories in solenoidal magnetic fields. Similar effects occur in other types of tracking problems, e.g., partially observed elliptical orbits in astrodynamics, etc., and are reminiscent of spherical aberration in ray tracking. Therefore an automatic method to handle the resulting nearly singular covariance structure warrants investigation. 
 
We make use of the generalized Cholesky decomposition of a particular {\em indefinite} work matrix, $\bf K$ ($k$ by $k$), in order to construct a restricted maximum likelihood function (ReML) via the method described in \cite{2001a}. This method is generalized to handle complex arithmetic in the indefinite case. We selected the ReML technique because our problem is well matched to minimizing residuals with respect to the fixed effects without also estimating those fixed effects. In our particular case the fixed effects correspond to the mean position of a vertex. ReML eliminates the fixed effects from the likelihood. As will be shown the leading $k-1$ diagonal elements can be matched with the expansion of the determinant which normalizes the ReML gaussian distribution and the $k$\,-th diagonal element can be matched to the $\chi^2$ argument of the exponential term. In addition, because the problem at hand involves a nearly singular covariance structure, the corresponding near-zero diagonal terms, as compared with a tunable cut parameter, of the same Cholesky decomposition designate those rows of $\bf K$ which fold into vectors which form auxiliary conditions such that the corresponding rows and columns of $\bf K$  drop out of the primary likelihood construction and are set aside to form a set of restrictions (constraints) in the maximum likelihood problem. 

The procedure we use is based on the Newton-Raphson method and requires fast computation of the first and second derivatives of the likelihood function to construct the linear system for Newton-Raphson and to proceed through enough iterations before convergence is achieved. The calculation of these derivatives requires computation of the first and second derivatives of the Cholesky decomposition as well as the auxiliary conditions (as well as Lagrange multipliers depending on the details of the implementation). In order to speed up the numerical calculations we incorporated the method of backwards differentiation. The advantage of
backward differentiation is that the gradient vector (m by 1) can be calculated in one pass through function evaluation, and also the Hessian (m by m) in m passes through function
evaluation. Also, backward differentiation of the Cholesky algorithm comes with a greatly reduced computer memory requirement.

\cite{Smith1995} describes how to efficiently compute both forward and backward derivatives of the Cholesky decomposition by using methods taken from automatic differentiation \citep{Griewank2000}. This technique permits variance-covariance estimation by restricted maximum likelihood (ReML). While the Cholesky decomposition and its derivatives are finding applications with ReML and in linear models typical to studies of animal breeding \citep{MeyerandSmith1995}, these are coming with additional innovations (Meyer, 2001; Meyer and Kirkpatrick, 2005).\  
The computational methods have been introduced in general statistical packages, including SAS (SAS Institute, 2009; Meyer, 2007).

Outside of ReML, the Cholesky decomposition and its derivatives are finding the following applications: in spatial modeling or kriging \citep{Toal 2009}; in
lattice models with an example showing non-parametric curve fitting and cross-validation \citep{Smith 1997}; in optimization involving the inverse diffusion problem \citep{Christianson 1997}; to differentiate a Laplace approximation of a likelihood function, thereby permitting estimations of parameters in a population model \citep{Frimannslund and Skaug 2006}; to permit Hamiltonian Markov Chain Monte Carlo in the context of non-linear regression by function factorization 
\citep{Schmidt 2009}; in calculating price sensitivities associated with exposure risk of financial portfolios \citep{Capriotti and Giles 2010};  and for optimizing several hyper-parameters within a gradient-based machine learning algorithm \citep{Bengio 2000}.  \citep{DeHoog 2011}\ have introduced a general notation for treating that task involved with calculating derivatives of functions that depend on triangular matrix factors (including Cholesky's factor), and they describe several application areas too.
 
The prior applications consider only the Cholesky decomposition of a positive-definite (or semi-definite) matrix. This convention is not followed in (Smith, 2001a and 2001b) 
where consideration is given to a linear state-space model, nor is it followed in the present paper. 
Smith used the Cholesky decomposition (and it derivatives) to estimate second moments, but in this different application the
Cholesky decomposition was applied to a symmetric and indefinite matrix. In general, the Cholesky decomposition may not be possible for an indefinite matrix, in which case an attempted decomposition may lead to an interruption. Such interruption is possible when a linear model includes effects that come with a singular
variance-covariance matrix structure as is possible with state-space models. It has been the case that an interrupted Cholesky decomposition permits the correct ReML likelihood calculation following 
(Smith, 2001a and 2001b).~
However, Smith's approach implies that the linear model is consistent with the singular variance structure, and in general this assumption cannot be made. The purpose of the present paper is to get beyond this limitation by introducing linear constraints that guarantee consistency, and hence this enlarges the set of cases where Cholesky-based ReML can be applied. There are also some minor typographical errors in the pseudocode listed in \citep{2001a}, and the corrected pseudocode is presented in the Appendix
(Section \ref{sec:pseudocode}) of this paper. Furthermore, it is the additional purpose of this paper to get beyond the common ReML applications and demonstrate a new Cholesky-based discrimination method that is suitable for enriching data samples that are used in the search for the Higgs Boson (``Higgs Hunting'') at the Large Hadron Collider (LHC).

A singular covariance matrix {\bf V} can arise in very predictable situations in linear models. 
For example, in multinomial or mixture problems the dependence in the
covariance matrix is well anticipated. Moreover, the coefficient matrix of the
normal equation may be singular because of over-parameterization, even if {\bf V} is
non-singular. However, the typical solution of striking out rows and columns of the
matrix {\bf V}, or of a coefficient matrix to the normal equations, does not work well in
the present situation where reliance on the Cholesky decomposition of an indefinite
matrix is made. Removing rows and columns of the indefinite matrix runs the risk of
removing too much information while ignoring computed constraints that come out
naturally from the Cholesky decomposition. The fact is that a singular error matrix
(that is also consistent with model specifications) implies that a linear
combination of the residuals either sum to zero or to a linear combination of the
fixed effects, and some model specifications (or parameters in the model) may be
inconsistent with computed constraints. The constraints must be checked to avoid
taking model consistency for granted, and this involves a completely new and
different treatment of singularities compared to past treatments.

The mixed linear model and ReML are reviewed in Section \ref{sec:mixed}. In Section \ref{sec:zero}, likelihood evaluation is cast in terms of residual error, and the Cholesky decomposition of a symmetric and indefinite matrix. Section \ref{sec:constraints}\ provides mathematical justifications for the constraints that may be needed when the Cholesky decomposition is interrupted. To treat the possibility of constraints, Section \ref{sec:Lagrange}\ describes optimization by Lagrange multipliers and Section \ref{sec:penalty}\ describes optimization by penalized maximum likelihood. Both methods are readily adapted to the Cholesky decomposition and its derivatives. Section \ref{sec:Siegel}\ treats estimation of fixed effects as an auxiliary calculation to the Cholesky decomposition by solving an indefinite system of equations. Section \ref{sec:vertex}\ presents an example coming from experimental physics, where the ReML method is used to devise a discrimination function to reduce backgrounds and preserve signal events in order to enrich data samples used for Higgs Hunting. Numerical stability and tuning of the algorithm is discussed in Section \ref{sec:Reliability} and the conclusion follows in Section \ref{sec:conclusions}.

\section{Mixed Linear Models, Log-Likelihood and ReML\label{sec:mixed}}

Notation: In the following Sections, we denote the transpose of a matrix $\bf R$ by $\bf R'$.
We also denote column vector in component form by square brackets, ${\bf v} = [a,b,c,..]$, and the corresponding row vectors with parentheses,
${\bf v'} =(a,b,c,...)$.

Linear models, including mixed linear models, have found widespread use in statistical investigations. The linear model, though additive, is frequently flexible enough for real situations as an approximation around the mean. Also, linear models and the associated normality assumptions are well understood. Methods as old as the analysis of variance (ANOVA) are completely consistent with mixed model methods. Furthermore, while the theory is developing in new areas, such the Gibbs Sampler or with other Bayesian methods, mixed model methods benefit as new tools come along. 
The mixed linear model is represented by:
$$\bf y=X\beta + Zu + \epsilon,$$where $\bf y$ is a vector of observations, $\bf \beta$ is a vector of fixed effects, $\bf u$ is vector of random effects and $\bf \epsilon$ is the observational error. The matrices $\bf X$ and $\bf Z$ are incidence matrices that relate the various effects to observations. The first moments for the random effects (their expectations) are $\rm E[{\bf u}] = 0$  and $\rm E[ {\bf \epsilon} ]=0$, and the variance-covariance structure is given by ${\rm var}[{\bf u}]=\bf G$, ${\rm var}[\epsilon]=\bf R$ and ${\rm cov}[{\bf u},{\bf \epsilon} ] =0$. Additional assumptions are needed to implement maximum likelihood or computer simulation, and generally $\bf y$, $\bf u$, and $\bf \epsilon$ are taken as multivariate normal. As indicated in \citep{Goldberger 1962}, 
the Best Linear Unbiased Prediction (BLUP) of $\bf u$ is found by evaluating 
\[ \bf \hat u = G Z'V^{-1}[y - X\hat\beta],\]
where \begin{equation}\bf V={\rm var}(y)= ZGZ' + R \label{eqn:V}\end{equation}
and where $\bf \hat\beta$ is the Best Linear Unbiased Estimate (BLUE) of the fixed effects obtained by the Generalized Least Squares (GLS) problem
\begin{equation} {\bf \left( X'V^{-1}X\right)[\hat \beta] = [X'V^{-1}y]}. \label{eqn.gse}\end{equation}
These equations can be reformulated so that the solutions can be obtained directly from the mixed model equations \citep{Henderson}
\begin{equation}
\bf \left(\begin{array}{cc}
\bf X'R^{-1}X & \bf X'R^{-1}Z  \\ 
\bf Z'R^{-1}X & \bf Z'R^{-1}Z + G^{-1} \\ 
\end{array} \right)
\left(\begin{array}{c}
\bf \hat\beta \\
\bf \hat u\\ 
\end{array} \right) = 
\left( \begin{array}{c}
\bf X'R^{-1}y \\
\bf Z'R^{-1} y\\
\end{array}\right).
\label{eqn:mixed}
\end{equation}
However, Section \ref{sec:Siegel}\ (see below) provides and alternative method based on the method of \citep{Siegel}.

It is well known that if a non-informative prior is used to describe the fixed effects in a Bayesian context, the posterior distribution (conditional on $\bf y$) of a linear combination of ${\bf b}$ (where ${\bf b} = [{\bf \beta', u'}]$), say $\bf Hb$, is multivariate normal. In this case the posterior distribution has mean vector $\bf H\hat b$ (where ${\bf \hat b} = [{\bf \hat\beta', \hat u'}]$) and variance-covariance matrix given by $\bf HC^{-1}H'$, where $\bf C$ is the 2-by-2 partitioned matrix on the Left Hand Side (LHS) of Eq. (\ref{eqn:mixed}). Therefore the mixed model equations are only good when the inverses $\bf R^{-1}$ and $\bf G^{-1}$ both exist. 
If either $\bf R^{-1}$ or $\bf G^{-1}$ does not exist, then $\bf C$ does not exist. Therefore, in the case where the  covariance structure is singular, the mixed model equations will not apply.

The log-likelihood for the Multivariate Normal (MN) is given by
$$ {\bf \ln(MN)} = {\rm constant} - {1\over2}\ln|{\bf V}| -{1\over2}\bf (y-X\beta)'V^{-1}(y-X\beta)$$
The maximum likelihood estimates of $\bf \beta$ and the dispersion parameters ($\bf R$ and $\bf G$) are found by maximizing the log-likelihood. Estimates of the dispersion parameters can be badly biased by small-sample errors induced by the estimation of $\bf \hat\beta$. This is a serious problem when the dimension of $\bf \beta$ is large relative to the information available to estimate $\bf \beta$.

To overcome this problem \citep{PandT}\ 
introduced Restricted Maximum Likelihood (ReML), where the dispersion parameters are found by maximizing 
\begin{equation}
{\bf \ln(ReML)} = {\rm constant} -{1\over2}\ln|{\bf V}| -{1\over2}\ln|{\bf X'VX}| - {1\over2}\bf (y-X\hat\beta)'V^{-1}(y-X\hat\beta)\label{eqn:ReML},\end{equation}where $\bf \hat\beta$ is the solution obtained by GLS, Eq. (\ref{eqn.gse}). 
ReML has the advantage of estimating away the $\bf \beta$ parameters from the likelihood. This is especially useful in cases where one wants to concentrate on minimizing the deviations from a common mean, without explicitly finding that common mean. \citep{Wiki}\ states, ``In contrast to conventional maximum likelihood estimation, ReML can produce unbiased estimates of variance and covariance parameters.'' \citep{Harville1}\ 
derived the likelihood in Eq. (\ref{eqn:ReML}) to treat the ``error contrasts'' which are found by taking a complete set of linear combinations of the observations which are sufficient to remove the effect of $\bf \beta$ while leaving the maximal amount of information for the purpose of ReML. An early review of ReML can be found in \citep{Harville2}. 
More reviews can be found in (Speed, 1977 and 1995).~
The relevance for the particular application described in Section \ref{sec:vertex}\ (see below), is that the $\chi^2$ that is determined by ReML is independent of the central vertex coordinates. It is possible to back-out the coordinates of the fitted vertex, but it is not necessary to know the coordinates in order to determine the goodness of fit. 
  
\section{Applying ReML to the $\bf Z = 0$ Scenario\label{sec:zero}}
 
Consider the following linear model $\bf {\cal L}$ in the case where $\bf Z = 0$
$$\bf {\cal L}: y = X \beta + \epsilon,$$
where $\bf y$ is an independent observation vector, $\bf \beta$ is a vector of fixed effects which are to be removed using the ReML likelihood, and $\bf \epsilon$ is a vector of random residuals. 
The array $\bf X$ is the incidence matrix that assigns fixed effects to observations. 
The variance-covariance matrix of the random residuals is denoted by $\bf R$ $${\rm var}[{\bf \epsilon}] = \bf R.$$ 
The linear model $\bf \cal L$ is now fully specified and the error matrix satisfies $\bf V = R$ so that the likelihood 
given by Eq. (\ref{eqn:ReML}) reduces to
\begin{equation}\bf \ln(ReML) \rightarrow  {\rm constant} -{1\over2}\ln|R| -{1\over2}\ln|X'RX| - {1\over2}(y-X\hat\beta)'R^{-1}(y-X\hat\beta).\label{eqn:disc}\end{equation}

\citep{2001b}\ 
associates a symmetric matrix $\bf K$ with the above linear model $\bf {\cal L}$ as follows:
\begin{equation}\bf K = \left( \begin{array}{ccc}
\bf R & \bf X & \bf y \\
\bf X' & 0 & 0 \\ 
\bf y' & 0 & 0\\
\end{array} \right),  \label{eqn:K}\end{equation}where the 0's denote the appropriate-sized null square matrices required to fill out the rows and columns of $\bf K$ and $\bf X'$ is the transpose of $\bf X$.  
Therefore, $\cal L$ implies the existence of a matrix $\bf K$, and $\bf K$ implies there exists a model $\cal L$.
Our main interest in Eq. (\ref{eqn:K}) is that $\bf R$ is {\it not required} to be invertible. The method we use is able to identify those linear combination of $\bf y$ which are associated with the zero-eigenvalues of $\bf R$. These linear combinations are non-stochastic and can be eliminated from the stochastic part of the likelihood and treated by the method of Lagrange constraints. In other words, our method is able to find the natural constraints required for the maximum likelihood problem for ReML. 

It is well known that the Cholesky decomposition runs to completion with any matrix that is symmetric and non-negative definite. \citep{2001b}\ 
shows that the Cholesky decomposition can
also be performed on the matrix $\bf K$, and this is most curious because while $\bf K$ is symmetric, it is not non-negative definite. $\bf K$ is classified as indefinite. 
However, the rows and columns of $\bf K$ must be first permuted, leaving the last row and column in place as required. 
This is enough to permit computation of the Cholesky decomposition, or the lower triangular matrix $\bf L$, where $\bf K_{k \times k} = LL'$ (for some permutation involving the first $k-1$ rows and columns). 
(Smith, 2000 and 2001a)~
describes the ReML function, or the likelihood that removes the impacts of  $\bf \beta$, to be a function of this particular $\bf L$. 
This calculation is given neatly as follows:
$${\bf \ln(ReML)} \rightarrow {\rm constant} - \sum_{i<k} \ln|L_{i,i}| + {1\over2}L^2_{k,k},$$where the absolute modulus function $|L_{i,i}|$ transforms a possible imaginary number into a positive number, and the summation is only over the non-zero pivots. 
When zero pivots are encountered (when $L_{i,i}=0$ for some $i$) {\it we require that the $i$-th column of $\bf L$ vanishes and this is enough to justify likelihood evaluation by the above formula}\/. 
It is easy to show that this calculation agrees with the likelihood given in Eq. (\ref{eqn:disc}) when R is non-singular.
However, in the most general case we cannot simply skip the zero pivots, and more will be said about this in the Section \ref{sec:constraints}, because particular constraints must be imposed.

The likelihood function is derived from elements of the Cholesky decomposition, and so there is nothing else that is needed to perform ReML but to find the derivatives that permit the likelihood to be optimized by the iterative Newton-Raphson technique. These derivatives come automatically with the Cholesky decomposition 
(Smith, 2000 and 2001a),~
and so there is little beyond the matrix $\bf K$, and its decomposition, that must be considered to describe ReML. The corrected pseudocode for the differentiation of the Cholesky algorithm and directions for how to use it are found in the Appendix (see Section \ref{sec:pseudocode}).

\section{Justification and Additional Constraints \label{sec:constraints}}

The Cholesky decomposition of a matrix $\bf K_{k\times k}$ proceeds by identifying a non-zero diagonal element (the first pivot), 
and then permuting rows and columns of $\bf K$ to reposition that diagonal to the first diagonal position. 
Elementary row operations are now conducted to annihilate elements below the diagonal in the first column (this is pivoting).
The first row of $\bf K$ is now transformed into the first column of $\bf L$ by replacing the first diagonal by its square root
and by dividing the remaining elements in the first row by that square root. 
The first row and column of $\bf K$ are now deleted to produce a smaller sub-matrix of order $k-1$. 
This sub-matrix is called the Schur complement. An outline computation of the first Schur complement (of $K_{11}$ in $\bf K$) is displayed below.
\[  
\left( \begin{array}{cccc}
K_{11} & K_{12} &  ...  & K_{1k} \\
K_{21} & K_{22} &  ...  & K_{2k} \\
   ...      &    ...      &   ...  & ..        \\
K_{1k} & K_{k2} &  ...   & K_{kk} \\  
\end{array} \right) 
\rightarrow
\left( \begin{array}{cccc}
K_{22} & K_{23} & ...  & K_{2k} \\
K_{32} & K_{33} & ...  & K_{3k} \\
    ...      &    ...    &  ...  & ...      \\
K_{k2} & K_{k3} & ...   & K_{kk} \\  
\end{array} \right) 
- {1\over K_{11}}
\left( \begin{array}{c}
K_{21} \\
K_{31} \\
    ...     \\
K_{k1} \\  
\end{array} \right) (K_{12}, K_{13},  ... , K_{1k})
\]
When $\bf K$ is symmetric, the Schur complement is symmetric. 
Therefore, the Cholesky decomposition may proceed by half-storing $\bf K$. 
The above steps are now repeated to generate a second Schur complement, then a third and so on.
The algorithm may be organized to overwrite the half-stored $\bf K$ with $\bf L$. 
The operations of the Cholesky decomposition are reversible, and therefore, the Cholesky decomposition conserves information.

\citep{2001b}\
generalized the Cholesky decomposition for the case when $\bf K$ is symmetric and indefinite by using a complex representation for the Cholesky diagonals if needed.
The sub-matrix initially containing the zero entries is the non-positive definite partition, 
and with fill-in (generated from pivoting from the diagonals of $\bf R$) the non-positive definite partition becomes more negative. 
When pivoting switches over to the diagonals of the non-positive definite partition, 
then fill-in results in the partition initially set to $\bf R$ (or the non-negative definite partition) thereby making it more positive. 
If a few pivots are selected from the diagonals of the non-negative definite partition first, before switching over to the 
non-positive partition and continuing pivoting over the negative diagonals until the non-positive definite partition returns 
to a matrix containing zero in all its entries (expect the last diagonal), then \citep{2001b}\
referred to the particular pivot order as a {\it standard data reduction.}
							
After the first pivot step within the Cholesky decomposition, the lead row and column is removed from $\bf K$, and this reduces the dimension of the resulting Schur complement by 1 as noted above. 
After a standard data reduction, the Schur complement is again reduced in dimension from $\bf K$ but retains the special matrix form (ignoring the last diagonal with no loss in generality):
\[ \bf  K_1 = \left( \begin{array}{ccc}
\bf R_1 & \bf X_1 & \bf y_1 \\
\bf X'_1 & 0 & 0 \\
\bf y'_1 & 0 & 0\\
\end{array} \right),  \]
where the subscript indicates that $\bf K_1$ was generated from $\bf K$ following the Cholesky decomposition. 
We will denote $\bf K$ as $\bf K_0$, to maintain consistency. 
In shorthand, this transformation is denoted by $\bf K_0 \rightarrow K_1$.

\citep{2001b}\  
notes that $\bf K_1$ also represents a model ${\cal L}_1$, given by 
$$\bf {\cal L}_1: y_1=X_1\beta_1 + \epsilon_1,$$
where $\bf \beta_1$ is sub-vector of $\bf \beta$ and the variance-covariance matrix of $\bf \epsilon_1$ is $\bf R_1$. 
The model ${\cal L}_1$ is fully determined given the Schur complement $\bf K_1$. 
Moreover, those observations that have been processed and eliminated from $\bf K$ are uncorrelated with $\bf \epsilon_1$.
Also, the extracted information was accounted for in the previously constructed Cholesky diagonal elements. 
In other words, the standard data reduction separates the data into two statistically independent parts. 
The first statistically independent part is used to compute that block of the log-likelihood already known for ReML and consistent with Eq. (\ref{eqn:disc}). 
The remaining part pertains to the treatment of ${\cal L}_1$, but because $\bf K_1$ is in the form of $\bf K$, the process can be iterated and the Cholesky decomposition continued to the next pivot.  

The Cholesky decomposition describes a sequence of standard data reductions:
$$\bf K_0  \rightarrow K_1 \rightarrow K_2 \rightarrow K_3 \rightarrow K_4\rightarrow ... $$ 
And these Schur complements correspond to a sequence of models:
$${\cal L}_0  \rightarrow {\cal L}_1 \rightarrow  {\cal L}_2 \rightarrow  {\cal L}_3 \rightarrow {\cal L}_4\rightarrow ... $$ 
At each transition the dimensions of the Schur complement get smaller and smaller, and all along the way information is being processed correctly to evaluate the ReML likelihood. The only question is whether the Cholesky decomposition completes and
leads to a final Schur complement that folds into the likelihood function calculation. 
However, the last Schur complement may be one of two special forms which cannot be
reduced further. 
In this case, the last Schur complement corresponds to the last model 
${\cal L}$ which is necessarily non-stochastic 
(i.e., the residual vector in ${\cal L}$ vanishes because its corresponding covariance matrix is computed to be null),  but consistent with the model 
specifications that are implied by a singular variance structure.
This last model is
contained in a Schur complement of the First Special Form: 
\[  \left(\begin{array}{ccc}
\bf 0 & \bf H & \bf v \\
\bf H' & \bf 0 & \bf 0 \\
\bf v' & \bf 0 & \bf p\\
\end{array} \right),  \]or the Second Special Form:
\[  \left( \begin{array}{cc}
\bf 0 & \bf v \\
\bf v' & \bf p \\
\end{array} \right),  \]
In the case of an incomplete Cholesky decomposition, what is made manifest are the constraints that must also be imposed on the maximization of the likelihood function. The remarkable conclusion is that all the information is contained within the Cholesky decomposition (as we will see), even when the Cholesky decomposition is unable to finish.

One might question the appropriateness of the standard data reduction, given that the pivot order in the Cholesky decomposition may be dynamic and may not follow the standard data reduction. 
However, \citep{2001b}\  
proved that pivot orders come in equivalence classes. 
Any dynamic order corresponds to a sequence of standard data reductions where the ReML likelihood is treated correctly. 
And moreover, Schur complements are invariant to the pivot order that generates them (including the last diagonal of the Cholesky decomposition), 
as well as the determinant calculated as the product of pivots. 
The correct likelihood is calculated even when the standard data reduction is not followed, because implicit in any pivot order is a sequence of standard data reductions.

With dynamical pivoting, the Cholesky decomposition is permitted to go as far as it can while skipping zero diagonals that would otherwise be pivots. 
Some zero pivots may encounter fill-in during the computation, and this permits the Cholesky decomposition to continue with additional rounds of pivoting. 
If for some reason the Cholesky decomposition in unable to complete the operations, the matrix that remains (as unfinished) will be a sub-matrix of what had been $\bf K$ and what was becoming $\bf L$ (but never completed).  
The unfinished sub-matrix contains the last Schur complement either in the First or Second Special Forms (noted above), 
even if the pivot order did not follow a particular sequence of standard data reductions.

If extraneous parameters remain impacting the likelihood function, they are found involved in non-stochastic linear combinations (indicated by the last model ${\cal L}$), 
and these combinations are revealed in the Schur complement (the First Special Form containing both $\bf H$ and $\bf v$) 
that could not be reduced by the Cholesky decomposition: 
$$\bf H\beta_I=v,$$where $\bf \beta_I$ is a sub-vector of $\bf \beta$, and $\bf v$ is a revealed linear combination of $\bf y$.  
If the extraneous parameters are no longer present, then the matrix $\bf H$ goes away (in the last model ${\cal L}$). 
We are left with the Second Special Form of the Schur complement that only involves $\bf v$ equated to a column vector of zeros: $$\bf v=0.$$ 
When the Cholesky decomposition is unable to finish, then one of these systems of linear equations becomes a side condition (a constraint) for the likelihood maximization exercise.

We may be uninterested in the extraneous parameters. 
However, we are interested in consistent models that don't contradict themselves when linear combinations are made of their components.
This concern is quite independent of the extraneous parameters as we will see.
The set of non-stochastic equations that are revealed (by the First Special Form of the Schur complement) 
will involve extraneous parameters, and these will be ignored in as much as ReML removes their impacts from the likelihood. 
The revealed set of equations can imply whatever they want about $\bf \beta_I$ and ReML will ignore them. 
This assumes that the statistical model is already consistent. 
However, the non-extraneous parameters that are identified for likelihood evaluation need not 
conform to a consistent linear model and that's where we depart from \citep{2001b}.

Likewise, if the attempt at Cholesky decomposition ends with a Schur Complement of the Second Special Form 
(leading to the equation $\bf v=0$), what is revealed are constraints that must be imposed to guarantee a consistent linear model. 
In this case there are no extraneous parameters to confuse the issue.

\section{Constraints: Lagrange Multiplier Method\label{sec:Lagrange}}
In the event that the Cholesky decomposition encounters a zero and is unable to complete,  in order to to maintain consistency one can consider appending to the ReML likelihood a Lagrange multiplier expression according to one for the following cases:

$${\rm Case~1:}~~~  F_1({\bf L},\lambda) = {\rm constant} - \sum_{i<k} \ln|L_{i,i}| + {1\over2}L^2_{k,k} + {\bf \lambda' (H\beta_I - v)},$$
where maximization proceeds with respect to the non-extraneous parameters $s_1,s_2,s_3,s_4$ (see Section \ref{sec:vertex}), $\bf \beta_I$ and $\bf \lambda$. 
Therefore, the side condition enforces the restriction that $\bf v$ is in the column space of $\bf H$ independent of $\bf \beta_I$.
	
$$ {\rm Case~2:}~~~ F_2({\bf L},\lambda) =  {\rm constant} - \sum_{i<k} \ln|L_{i,i}| + {1\over2}L^2_{k,k} + {\bf \lambda' v},$$			
where maximization proceeds with respect to the non-extraneous parameters $s_1,s_2,s_3,s_4$ (see Section \ref{sec:vertex}) and $\bf \lambda$. 
Therefore, the side conditions enforces the restriction that $\bf v=0$.

In both cases, we find an objective function so constructed from the known elements of the unfinished Cholesky decomposition. 
First and second derivatives are available \citep{2001a}\ 
and constrained optimization is straightforward by the technique of Newton-Raphson iteration as applied in Section \ref{sec:vertex}\ below.  

In paradoxical cases, the data may produce an inconsistent model. In which case, we recommend constrained maximization as noted above, and in this way useful information is included that would otherwise be ignored.

In case the Cholesky decomposition is unable to finish and the Schur complement is of the Second Special Form, then the particular linear combination of elements corresponding to $\bf v$ can be constrained to zero by adding a Lagrange multiplier term $\lambda' \bf v$ as indicated above. An alternative method involves adding a quadratic penalty term to the likelihood (see Section \ref{sec:penalty}) can be used to adjust the constraints to zero within estimated precision.

Case 1 and Case 2 above represent possible objective functions. However, we concentrate on Case 2 because that is the one relevant to our example. The objective function to be maximized is of the form $F_2({\bf L},\lambda)$. The three algorithms of Section \ref{sec:pseudocode}\ are used to construct the Newton-Raphson linear system. If required, the first derivatives with respect to the Lagrange multipliers (the $\lambda$ vector) come directly out of the $\bf v$ elements of the corresponding Schur complement of the unfinished matrix $\bf L$. Table 1 sketches the Cholesky decomposition which is used to calculate the components of the ReML likelihood and the possible non-stochastic linear combinations that will be subject to the constraint conditions. We initialize the array $\bf F$ of Table 2 to the array $\partial F_2({\bf L},\lambda)/\partial L_{ij}$ that represents the constrained objective function including the Lagrange multipliers. We then find the mixed second derivatives directly from $\bf Q$ (see Table 3, which represents the forward derivative calculations.)

\section{Constraints: Penalized Likelihood Method\label{sec:penalty}}

There is another method of implementing constraints besides Lagrange multipliers. Instead of introducing a set of Lagrange multipliers which are treated as independent parameters to be varied in a maximum likelihood procedure, the terms which they multiply can be squared and then multiplied by a fixed constant. This is known as the ``Penalized Likelihood" method. In the Penalized Likelihood method the Lagrange equations for Case 1 and Case 2 above are modified to the form 

$${\rm Case~1:}~~~  F_1({\bf L}) = {\rm constant} - \sum_{i<k} \ln|L_{i,i}| + {1\over2}L^2_{k,k} + {\bf \sum c_i (H\beta_I - v)_i^2},$$where the $c_i$ are nominated constants and maximization proceeds with with respect to the non-extraneous parameters $(s_1,s_2,s_3,s_4)$  and $\bf \beta_I$\ (see Section \ref{sec:vertex}). 
	
$$ {\rm Case~2:}~~~ F_2({\bf L}) =  {\rm constant} - \sum_{i<k} \ln|L_{i,i}| + {1\over2}L^2_{k,k} + {\bf \sum c_i(v_i)^2},$$where $c_i$ are nominated constants and maximization proceeds with respect to the non-extraneous parameters $s_1,s_2,s_3,s_4$ (see Section \ref{sec:vertex}). The squares of the column vectors in the last term on the RHS of the above are to be understood as vector dot products. 

The main difference between the Lagrange Multiplier method and the Penalized Likelihood method is that the auxiliary constraints are imposed to the maximum machine precision by the Lagrange Multiplier method, whereas the constraints are imposed in a less extreme manner with the Penalized Likelihood method. The point of the nominated constants is to adjust the enforcement of the constraints to within reasonable limits set by the known precision of the measurements. This allows the possibility, for example, of not demanding that the constraints be satisfied with more rigor than the measurement uncertainties can justify.

\section{Estimating the Fixed Effects $\bf \hat\beta$\label{sec:Siegel}}

\citep{Siegel}\ described a method to compute generalized least squares (GLS) estimates by way of a system of equations with a symmetric and indefinite coefficient matrix. 
In our notation, this system of equations is presented below:
\[
\left( \begin{array}{cc}
\bf R & \bf X \\
\bf X' & \bf 0 \\
\end{array} \right)
\left( \begin{array}{c}
\bf \lambda \\
\bf \hat\beta \\
\end{array} \right)=
\left( \begin{array}{c}
\bf y\\
\bf 0 \\
\end{array} \right), \]where $\bf \hat\beta$ is the GLS estimate of $\bf \beta$, and $\bf \lambda$ are the Lagrange multipliers. 
This Section will follow \citep{2001a}, and show how to calculate both $\bf \hat\beta$ and $\bf \lambda$ as adjunct operations that depict backward substitution, 
given that the Cholesky decomposition of $\bf K$ is available. 
 
Note that the both the coefficient matrix and right-hand side of Siegel's equations are sub-matrices of $\bf K$: the coefficient matrix is the lead sub-matrix in $\bf K$, 
and the right-hand side is fixed to the last column (or row) and remains there for all permitted row-column permutations of $\bf K$.

Now rewrite Siegel's equations in simple terms: $$\bf C b = r,$$ where $\bf C$ and $\bf r$ signify the coefficient matrix and right-hand side, 
respectively; and $\bf b$ is a column vector containing $\bf \lambda$ and $\bf \hat\beta$. 
To solve these equations, we might permute the rows and columns of $\bf C$, and compute the Cholesky decomposition: 
$\bf TT'=C$. 
To permute the rows and columns of $\bf C$, and the rows of $\bf b$ and $\bf r$ are also permuted to leave Siegel's equations intact. 
Now multiplying both sides of Siegel's equations by the same elementary row operations that transform $\bf C$ into $\bf T'$ gives:
$$\bf T'b = a.$$When $\bf C$ is non-singular, $\bf a = T^{-1}r$. 
The coefficient matrix in this new system of equations is upper triangular. 
Therefore, vector $\bf b$ can be solved by backward substitution.

With $\bf L$ computed, where $\bf K= LL'$, note that $\bf T$ is the leading sub-matrix in $\bf L$ 
and $\bf a'$ is the last row vector of $\bf L$ (excluding the last diagonal). 
Therefore, having computed $\bf L$ we need only enter backward substitution to evaluate $\bf b$.  
The GLS estimates $\bf \hat\beta$ will be found scattered in $\bf b$, noting that $\bf b$ is permuted.

When the $i$-th pivot encountered in $\bf L$ is zero, and the $i$-th column vanishes, a singularity is present and $\bf b$ is not estimated uniquely. 
We place a restriction on $\bf b$: set its $i$-th entry to zero. 
This modification is implemented when the Cholesky decomposition is unable to finish but ends with a Schur complement of the Second Special Form. 
The associated column of $\bf L$ will also be constrained to vanish.
When the Cholesky decomposition ends with the First Special Form then an auxiliary estimate of $\bf \beta_I$ is available from the main optimization 
and this estimate is used in backward substitution given that what had been computed for the rest of $\bf L$ is lower triangular. 

\section{Example: Spatial Errors in Track Extrapolations and Vertexing in $\bf Higgs \rightarrow 4~lepton$ Searches\label{sec:vertex}}
\begin{figure}[hptb]
\begin{center}
 \resizebox{0.6\linewidth}{0.6\linewidth}{\includegraphics{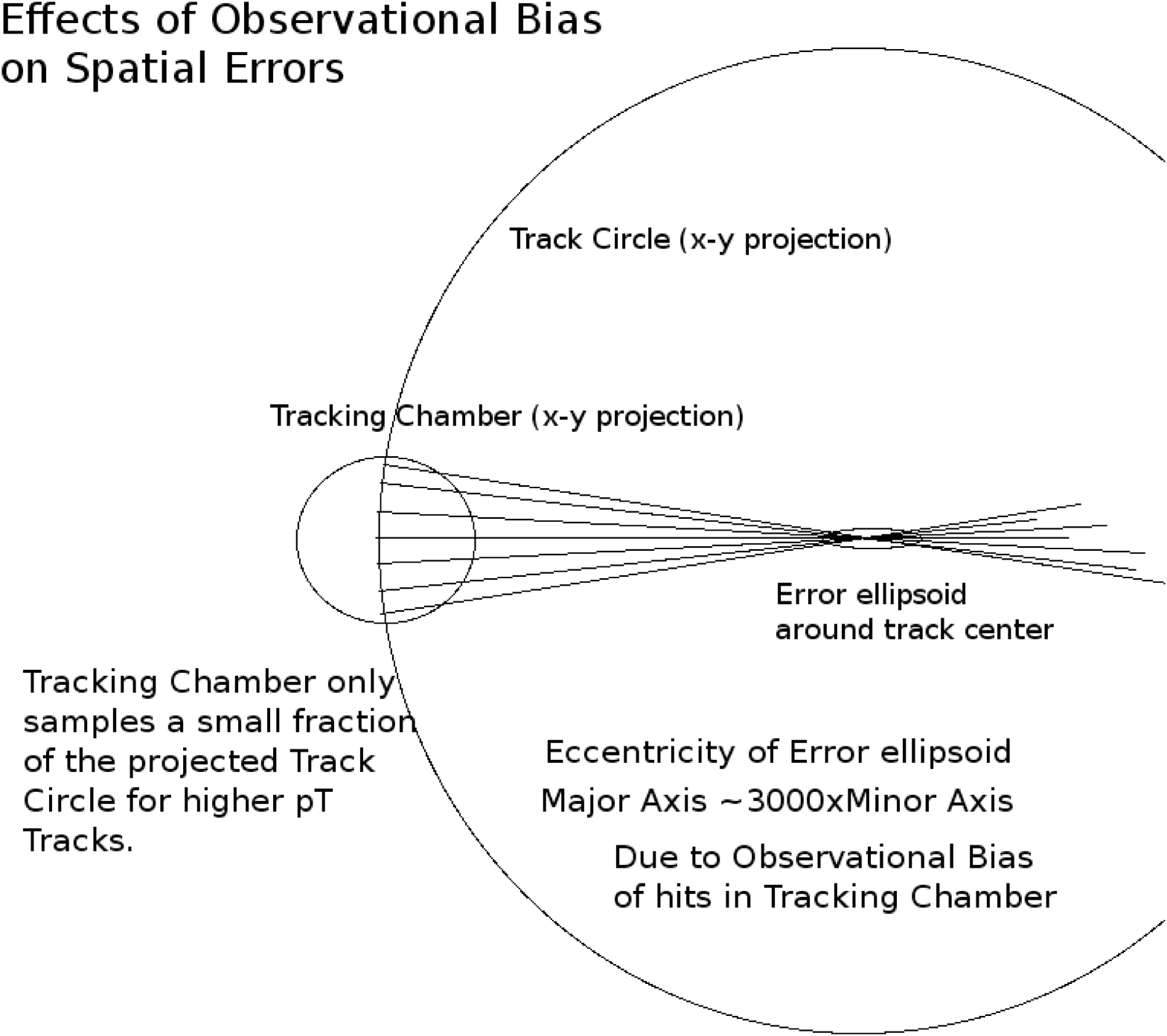}}
 \caption{Relation of Error Ellipsoid of Center of Curvature and Tracking Hits.}
 \label{fig:error_ellipsoid}
\end{center}
\end{figure}

One of the most important decay modes of the Higgs Boson is into 4 charged leptons $\rm (Higgs \rightarrow 4~leptons)$. In a solenoidal detector there is a large magnetic field present which is represented as a constant vector, $\vec{\bf B}$, which is taken by convention to point in the $z$-direction. Each of the charged leptons will approximately trace out a right-circular (or left-circular) helix with symmetry axis also parallel to the $z$-direction. Tracks undergo multiple scattering and energy losses as they traverse the detector which limit the accuracy of the helical path assumption. These effects are usually very small for tracks in the central solenoid region that have high transverse momentum $p_T$ (momentum component perpendicular to the $z$-axis). The radius of a track's helical path depends its $p_T$. Each helix in 3-d has the form $\vec r(s) = \left[x(s),y(s),z(s)\right]$ and is a function of an independent position parameter $s$ that marks the location along the track path. The components of the position vector $\vec r(s)$ are given by
\begin{equation}\begin{array}{ccl}
x(s) &=& (\rho-dxy)\sin\phi  - \rho\sin(\phi -{s/\rho}) \\
y(s) &=& \rho\cos(\phi -{s/\rho}) - (\rho-dxy)\cos\phi  \\
z(s) &=& Z  + s\cot\theta \\
\end{array}\label{eqn:coordinates}\end{equation}The parameters on the RHS of Eq. (\ref{eqn:coordinates}) are obtained from the track-fitting algorithm and have the following meaning:
\begin{itemize}
\item[]
\subitem{}$q = \rm charge~of~particle$
\subitem{}$k = \rm curvature~ of~ track~ circle, R = 1/k = \rm radius~ of~ curvature$
\subitem{}$Z =\rm ~\hbox{z-coordinate}~ of~ the~ point~ on~ the~ track~ helix~closest~ to~ \hbox{z-axis} $
\subitem{}$\theta  = \rm polar~ angle~ of~ the~tangent~to~track~at~ Z $
\subitem{}$\phi  = \rm azimuthal~ coordinate~ (\tan\phi =-x/y)~of~ the~ track~ helix~ at~ Z $
\subitem{}$\rho = qR$
\subitem{}$D  =\rm ~signed~ distance~ from~ beam~ axis~to~ track~ helix~ at~ Z $
\subitem{}$dxy = qD$
\subitem{}$dsz = Z/\sin(\theta)$
\end{itemize}
where the sign of the distance for $D$ is given as positive if the track circle (projection of helix into the $x$-$y$ plane) contains the z-axis and negative otherwise. A typical track fitting program will produce the 5-parameters $(\rho, \phi, \theta, dxy, dsz)$ along with a 5-by-5 covariance matrix which can be used to estimate the spatial error matrix in terms of the 3-by-3 correlation matrix for $x(s), y(s), z(s)$, for example, by the method of ``propagation of errors''. 
Once these track parameters are measured, then the track can be extrapolated to any position along its trajectory by using the parametric equation of a helix as a function of $s$. One of the interesting properties of these spatial extrapolations is the the 3-by-3 spatial error matrix is very close to rank-2 (almost singular). Fig. (\ref{fig:error_ellipsoid}) illustrates the reason that the spatial error matrix has a near-zero eigenvector lying almost entirely in the $x$-$y$ plane. If the radius of the track circle is much larger than the radius of the tracking detector, then only a small fraction of the circumference of the track is measured. This creates a measurement or observational bias in the sample of hits along the track. 
Such an observational bias creates a highly eccentric error matrix for the reconstructed $x$ and $y$ coordinates of the center-of-curvature. This highly eccentric error matrix produces a spatial error matrix along the track helix which has a very small eigenvalue in the direction of the projection of the track tangent vector in the $x$-$y$ plane. This property is independent of the method used to determine the track parameters precisely because the tracks we are most interested in have high $p_T$ and, therefore, have very large diameter track circles compared to the diameter of the tracking chamber. 

In order to determine if a Higgs candidate has been selected in the data-analysis of the experiment, one of the criteria applied to the 4 selected tracks is: Are these 4 tracks consistent with originating at a common position in space? Do these 4-tracks form a ``vertex'' in space? If the hypothetical Higgs Boson decayed into 4 charged leptons, then each lepton would follow its own helical trajectory, 
but all 4 leptons would converge on a common point in space corresponding to the point of decay of the Higgs Boson. Reducible 4-lepton background events, would be inconsistent with originating from the same common location. Therefore constructing a likelihood for the hypothesis that the 4 tracks originate at the same location is a very useful algorithm to separate signal events (Higgs candidates) from reducible backgrounds.

The tracks are assumed to come from a common point by hypothesis and the likelihood is obtained by first constructing the $\bf K$ matrix of Eq. (\ref{eqn:K}) above with the ${\bf y}_i(s_i) = [x(s_i), y(s_i), z(s_i)]$ for coordinates $x(s_i) $, $y(s_i)$, and $z(s_i)$ for each track ($i$ is numbered 1 through 4) with its respective independent variable $s_1, s_2, s_3$ or $s_4$. The $\bf R$ matrix in Eq. (\ref{eqn:K}) is obtained by the ``propagation of errors'' method which involves Taylor expansion of the functions $x(s)$, $y(s)$ and $z(s)$ about the mean positions. The overall matrix is obtained by appending each of the contributions from each track to the form the assembly $\bf y = [y_1, y_2, y_3, y_4]$ and similarly for $\bf R$. The incidence matrix $\bf X$ is similarly constructed using ${\bf X_i}$= 3-by-3 unit matrices (for $i=1,2,3,4$) and stacking four such matrices one-above-the-other.
Given 3-by-3 unit incidence matrices for ${\bf X}_i= \bf I_3$ (for the i-th track), then ${\bf\beta}= \vec{r}_c$. The extraneous parameters $\bf\beta$ are the central position coordinates of the vertex which can be estimated using the method of Section  \ref{sec:Siegel}.
This relationship completes the definition of the Linear Model.

Since the extrapolated spatial error matrices have the nearly rank-2 property, it is necessary to utilize the methods described above to construct a consistent ReML for testing the hypothesis of the common spatial origin of the 4 selected leptons as well as automatically constructing the non-stochastic linear combinations which have to be constrained to zero (a la the Second Special Form mentioned above). The constraints are determined based on comparing the size of the respective diagonal element in the Cholesky decomposition to a tunable threshold value which is optimized for signal events. When such a zero is encountered, the location on the main diagonal is noted and stored for subsequent determination of the constraint itself. The ReML likelihood and collateral constraints are determined as outlined above and the system of first and second derivatives are calculated using the Tables 1, 2, and 3 of the appendix. A linear system of equations is then formed by Taylor expansion of the objective function and a step is taken by Newton-Raphson (NR) iteration to move towards the maximum likelihood position. The coordinate functions of Eq. (\ref{eqn:coordinates}), evaluated at $s=0$, produce the coordinates of the helix at closest approach to the beam-axis ($z$-axis). Since a real Higgs would be produced very close to the colliding beam axis, then $s_1, s_2, s_3, s_4$ should all be initialized to $0$ at the beginning of the first NR step for fast convergence. Alternatively, one could perform a binary search about $s=0$ (for each track) to determine appropriate initial values for the NR method. 

After convergence, the algorithm produces a set of four parameter values $s_1, s_2, s_3, s_4$ which give the location along each track such that the assumptions of the model are satisfied. The $\chi^2$ value is obtained as the square of the last element on the main diagonal of the resulting decomposed $\bf K$ matrix ($\chi^2 = -L_{k,k}^2$). This variable is only approximately $\chi^2$ coming with degrees of freedom that are underestimated (by the number of positive pivots minus the number of negative pivots plus 1, minus 4 from estimating the position variables $s_1, s_2, s_3, s_4$) and we use the empirical distribution for $\chi^2$ in applications. ReML and $\chi^2$ are invariant with respect to the coordinates of the central position (the vertex), because the impact of the central position was removed from the likelihood. The crucial assumption of the the model is that there be a common origination point for the four tracks \emph{somewhere}\/.
If the $\chi^2$ from the fit is small, the model is satisfied and the hypothesis that the 4-tracks originated from a common point is consistent with the data. This is what we expect for $\rm Higgs \rightarrow 4~leptons$. Reducible background events (such as $t\overline{t}, Zbb \rightarrow 4~{\rm leptons}$) where the 4-leptons do not originate from the same point will not fit the hypothesis of the model and will have large $\chi^2$ values.   

In order to implement the above algorithm a sample of events was generated using the PYTHIA \citep{Torbjorn 2006}\ Event Generation program. These events were then simulated using the LDT Monte Carlo Program \citep{Regler 2007}\ and \citep{Valentan 2011}\  to determine the detector response and to obtain fits to the track parameters and their covariance matrix. The analysis of the fitted tracks and their error matrices was done using the Rave/Vertigo dataharvesting and vertexing environment \citep{Waltenberger 2011}\ with the above ReML algorithm inserted internally in the package. 

\begin{figure}[h]
\begin{center}
 \resizebox{0.5\linewidth}{0.5\linewidth}{\includegraphics{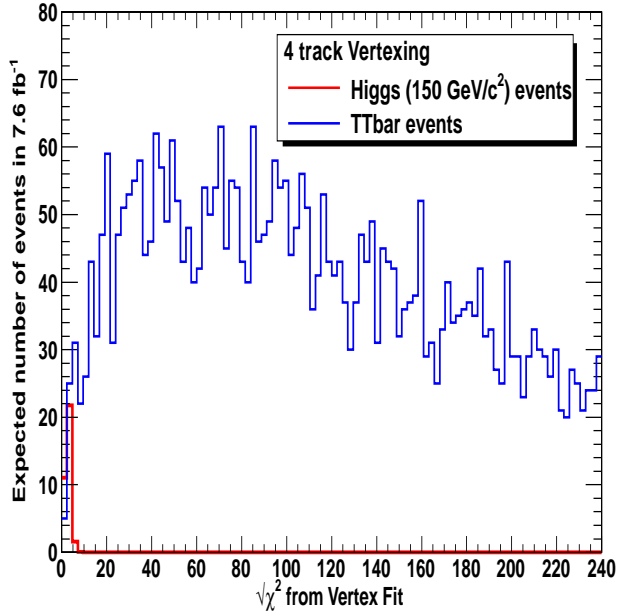}}
 \caption{Red histogram: Expected number of Higgs events at $\cal{L} =\rm 7.62~femptobarn^{-1}$ as a function of the ReML $\sqrt{\chi^2}$.
               Blue histogram: Expected number of $t\overline{t}$ background events at the same luminosity.}
 \label{fig:higgs}
\end{center}
\end{figure}

Applying the ReML method to the case of 4 leptons from Higgs decay and also from a known major source of background ($t\overline{t} \rightarrow 4~{\rm leptons}$), where the leptons can be either electrons or muons, we arrive at a comparison of signal and background shown in Fig. (\ref{fig:higgs}). Only events in which all four tracks had $p_T > 5 \rm ~GeV/c$ were used. 

\begin{figure}[h]
 \begin{center}$
 \begin{array}{cc}
\includegraphics[width=3.2in]{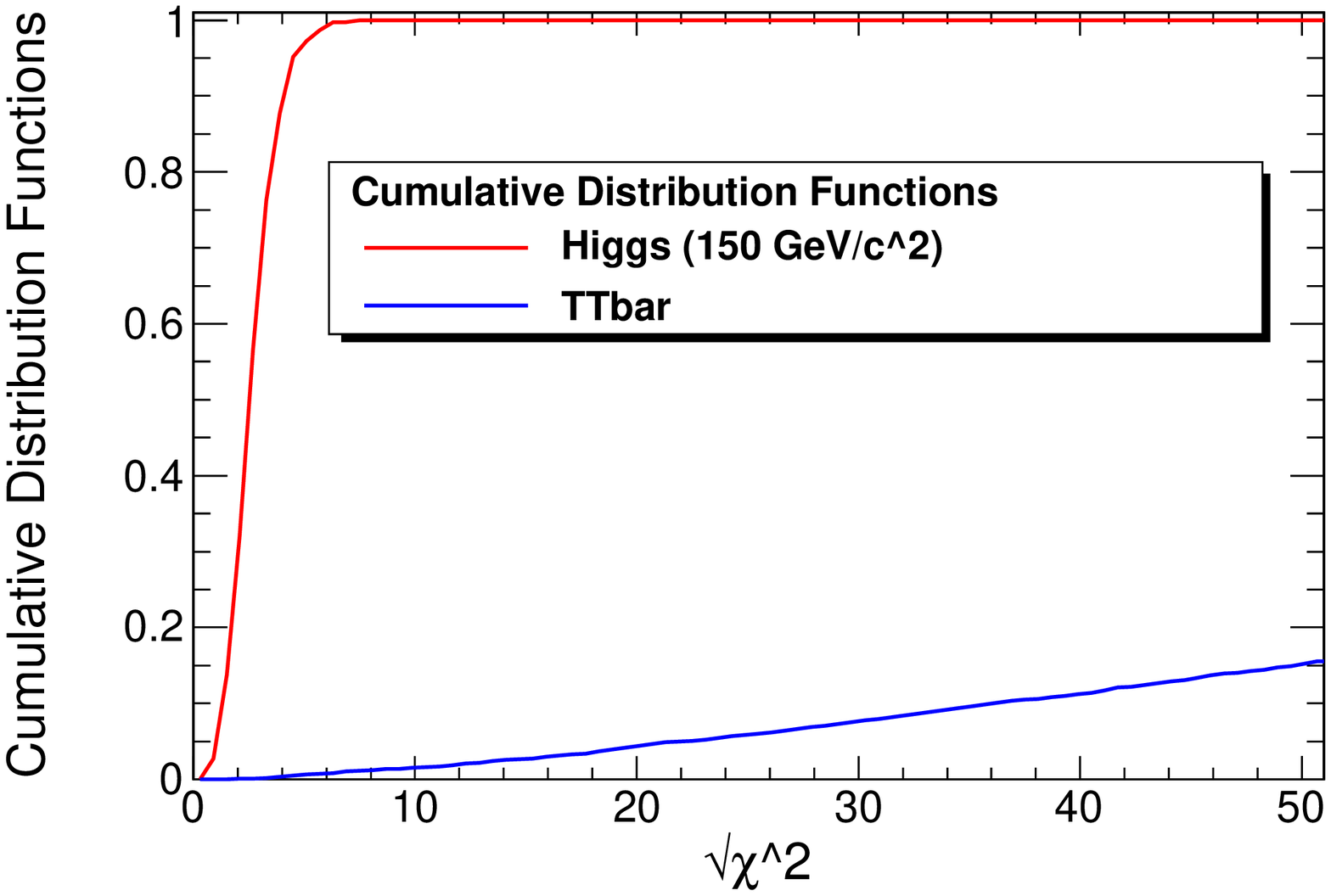} &
 \includegraphics[width=3.2in]{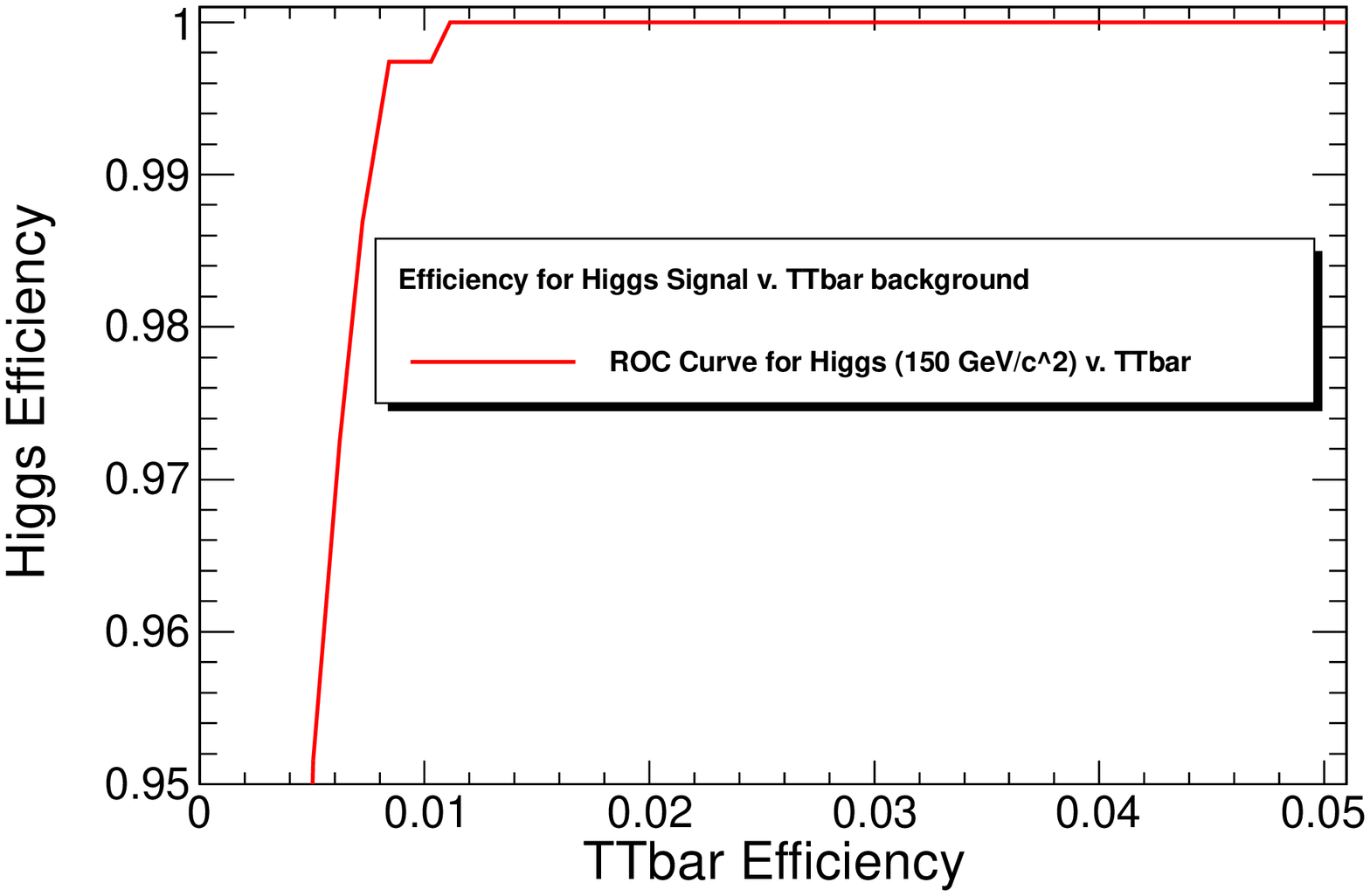}
   \end{array}$
     \caption{Left: Red histogram: Cumulative Distribution Function (CDF) for Higgs Boson events as a function of the ReML $\sqrt{\chi^2}$.
                  Blue histogram: CDF for  $t\overline{t}$ background events. 
                  Right: Cumulative distributions plotted against each other at the same $\sqrt{\chi^2}$ (Also known as Receiver Operating Characteristic or ``ROC'' curves). The CDF for accepting $\rm Higgs\rightarrow 4~lepton$ events is shown versus CDF for accepting $\rm t\overline{t}\rightarrow 4~lepton$ events based on the $\chi^2$ from 4-track vertexing. The step near the bend of the plot is due to fluctuations in the Monte Carlo statistics on the high end of the $\chi^2$ distribution in the signal events.}
   \label{fig:cdf}
 \end{center}
\end{figure}
  
As can be seen Fig. (\ref{fig:higgs}), even though the Higgs signal events have a smaller cross section and a smaller number of expected events than $t\overline{t}$ events, they are much more peaked near zero $\sqrt{\chi^2}$ for the same acceptance conditions of the detector. This results in a much higher efficiency (CDF value) for detecting Higgs as compared to retaining background events as can be seen in Fig. (\ref{fig:cdf}). This means that we can preferentially select Higgs candidates and reject $t\overline{t}$ background by cutting on the $\chi^2$ value.  

In the case of high luminosity collisions, there can be multiple events detected every time the detector is triggered. These multiple events are recorded by the detector in the same time window and are called ``pile up''. The ReML method, as well as other 4-track vertexing methods, have the advantage of being robust against the effects of high pile up since it does not depend on an auxiliary determination of the best vertex (the Primary Vertex) that matches with the 4-tracks used in the above analysis. Finding the correct Primary Vertex increases with difficulty as the proton-proton collision luminosity increases. Discrimination methods, such as b-tagging \citep{Tomalin 2008}, will be adversely affected as the number of pile up events increases. At the LHC design luminosity, approximately 200 pile up events per beam crossing are expected. The issue of finding the best Primary Vertex is irrelevant to the 4-track vertex method. If one of the 4-tracks does not match with the other 3 because it is from an un-related pile up event, then both the b-tagging algorithm and the 4-track vertex method will assign a large $\chi^2$ to that event and it will be rejected. However, if the wrong Primary Vertex is used as a reference for the b-tagging then all 4-tracks will register a large b-tagging discriminant value even though they might be matched to a common spatial position (i.e., have small $\chi^2$ from the vertex fit) by the 4-track vertex algorithm. These events should not be discarded just because the wrong and irrelevant Primary Vertex was used as a reference marker. If the 4-tracks have a small value of $\chi^2$ from the 4-track vertex method, then that alone is sufficient to decide if they should be further analyzed.  

\section{Numerical Stability and Tuning\label{sec:Reliability}}
The 4-track vertex method involves a tunable parameter $\alpha$, which determines if a pivot (diagonal element) is classified as a near-zero (constraint) or to be included in the likelihood construction. Ideally $\alpha$ corresponds to the zero-precision of the machine. During the Cholesky algorithm processing steps pivots are compared with $\alpha$. If the Cholesky pivot value is less than $\alpha$, then the corresponding row associated with that pivot is counted as a constraint row and marked for inclusion in the Lagrange multiplier (or penalty) terms. With $\alpha$ tuned to maximize the number of reconstructed 4-track vertices on a Monte Carlo signal sample, the algorithm identified 2 constraints per four-track event. Other methods of vertexing based on ``rules of thumb'' concepts in track extrapolation apply a constraint for every track leading to 4 constraints for a four-track combination (e.g., so-called ``working in the perpendicular plane'' for each track). If $\alpha$ is set too low, then the numerical instabilities associated with attempted inversion of a nearly singular covariance structure appear. If $\alpha$ is set too high, then more constraints than necessary are imposed and the constraints begin to involve the fixed effects for too large a number of identified constraints corresponding to too large a value for $\alpha$. In practice the distribution of pivot values for signal Monte Carlo events can be used to tune the size of alpha for good distinction between the near-zero and the non-zero pivots. Once tuned, the algorithm reconstructs all the vertices consistent with the hypothesis of the associated tracks originating from a point.

Also convergence performance of the Newton-Raphson iteration can be fortified by using binary search method to check the initialization of the algorithm. Unless tracks are constructed pathologically, or have very low values of $p_T$ and curl up in the detector (curlers), then all the vertices converge in less than 30 iteration steps and much fewer for the case of signal events. Failures to converge are associated with a set of tracks which are inconsistent with the hypothesis that they originate from a single point in space. In the above analysis tracks were restricted to the set with $p_T > 5 \rm ~GeV/c$  to avoid curlers.

\section{Conclusion\label{sec:conclusions}}
This paper presents an automated matrix procedure for constructing the ReML likelihood function and also to identify non-stochastic linear
combinations that represent quantities that must be constrained to zero in the event that the error matrix describing
the data is singular. The method is applied to the problem of determining how close four tracks from a $\rm Higgs \rightarrow 4~lepton$ decay approach a common position in space. This method can be used to discriminate between signal and background events at experiments at the LHC. It has an advantage of being independent of pile up which is not the case for b-tagging.  

\section{Acknowledgments\label{sec:acknow}}
We are especially thankful for the assistance of Nicola De Filippis, (Politecnico and INFN Bari, Italy), who incorporated the above method
in the software for the CMS Experiment. The analysis outside of the CMS environment was made possible by the work of Manfred Valentan (Institute of High Energy Physics, Austrian Academy of Sciences, Vienna, Austria) who wrote the interface to read events generated by PYTHIA and provided the LDT Monte Carlo program to simulate and reconstruct the tracks. Without the help and encouragement of Wolfgang Waltenberger (Institute of High Energy Physics, Austrian Academy of Sciences, Vienna, Austria), it would have been impossible to develop the code necessary to insert the ReML algorithm into the Rave/Vertigo analysis environment. We thank Andrey Korytov and Alexey Drozdetskiy (University of Florida, Gainesville, Florida, USA) for interesting discussions on how to compare signal and background events using the method described above on Monte Carlo samples. Thanks also go to Robert Alan Wolf (University of San Francisco Mathematics Department, San Francisco, California, USA) and Chris Freiling (California Sate University Mathematics Department, San Bernardino, California, USA ) for their mathematical insights into the geometrical problems encountered in this work.  
 




\bibliographystyle{elsarticle-num}





{}

\newpage

\section{Appendix: Pseudocode for Differentiation of Cholesky Decomposition\label{sec:pseudocode} }
This Section includes corrections (shown in red) to the pseudocode which was given in \citep{2001a}. Construction of the Cholesky decomposition for an indefinite matrix ${\bf K}_{N \times N}$ is summarized in the following table \\

\begin{tabular}{l}
Initializations:\\
$\bf L$ is lower triangular and $L_{ij} \leftarrow ij$-th element of ${\bf K}_{N \times N}(x_1, x_2, ..., x_p)$.\\
$\ominus_k =$ ``$+$'' if $k$-th diagonal is part of non-negative submatrix, ``$-$'' otherwise.\\
$\pm_k = -\ominus_k$.\\
\hline\\
Algorithm:\\
For $k = 1, ..., N$ do\\
if $|L_{kk}| \approx \rm zero$, check to see if remaining $L_{jk} \approx \rm zero$, $j = k+1, ..., N$ and skip $k$-th pivot.\\
Otherwise, $L_{kk} \leftarrow {\rm sqrt}[\ominus_k L_{kk}]$ and do: \\
$L_{jk} \leftarrow \ominus_k L_{jk}/L_{kk}$, for $j = k+1, ...,N$,\\
$L_{ij} \leftarrow L_{ij} \pm_k (L_{jk}\times L_{ik})$, for $j = k+1, ..., N$ \& $i = j, ..., N$.\\
end $k$\\
\hline\\
Table 1. Pseudocode for Cholesky Decomposition with Possible Negative Diagonals.\\
\end{tabular}

\vspace{0.50truein}
First derivatives, $\partial F({\bf L}) / \partial x_v$, are summarized in the following table

\begin{tabular}{l}
Initializations:\\
$\bf L$ is provided (see Table 1).\\
{${\bf F}_{N \times N}$ is a work space with elements $F_{ij}$, defined respectively for $i \ge j$.}\\
$\ominus_k =$ ``$+$'' if $k$-th diagonal is part of non-negative submatrix, ``$-$'' otherwise.\\
$\pm_k = -\ominus_k$.\\
$F_{ij} \leftarrow \partial F({\bf L})/\partial L_{ij}$\\
\hline\\
Algorithm:\\
(a) ${\bf F} \leftarrow T({\bf F})$ by following operations:\\
For $ k = N, ..., 1$ (N.B. Decreasing Order) do\\
if $|L_{kk}| > \rm zero$, then do:\\
$F_{ik} \leftarrow F_{ik} \pm_k (F_{ij} \times L_{jk})$, $F_{jk} \leftarrow F_{jk}  \pm_k (F_{ij} \times L_{ik})$, for  $j = k+1, ...,N$ \& $i = j, ..., N$\\
$F_{jk} \leftarrow \ominus_k F_{jk}/L_{kk}$,  $F_{kk} \leftarrow F_{kk} \pm_k (F_{jk} \times L_{jk})$, for  $j = k+1, ...,N$\\
$F_{kk} \leftarrow \ominus_k (1/2) F_{kk}/L_{kk}$\\
end $k$\\
\\
\textcolor{red}{(b)   $\partial F({\bf L})/\partial x_v = \sum_{i \ge j} F_{ij}\times \partial K_{ij}/\partial x_v, \quad v = 1, 2, ... p.$}\\
\\
\hline\\
Table 2. Pseudocode for Backward Differentiation of $F({\bf L})$.\\
\end{tabular}

\newpage

\vspace{0.4truein}
Second derivatives, $\partial^2 F({\bf L})/\partial x_v \partial x_u$, are summarized in the following table
\vspace{0.4truein}\\

\begin{tabular}{l}
Initializations:\\
$\bf L$ is provided (see Table 1).\\ 
$\ominus_k =$ ``$+$'' if $k$-th diagonal is part of non-negative submatrix, ``$-$'' otherwise.\\
$\pm_k = -\ominus_k$.\\
${\bf F} \leftarrow T(\partial F({\bf L})/\partial L_{ij})$ provided (see Table 2).\\
${\bf S}_{N \times N}$ and {${\bf Q}_{N \times N}$} are work spaces with elements $S_{ij}{, Q_{ij}}$, defined respectively for $i \ge j$.\\
$Q_{ij} \leftarrow \partial K_{ij}/\partial x_v,\quad i\ge j,\quad v\in \{1,2,..., p\}$.\\
\hline\\
Algorithm:\\
(a) For $k = 1,...,N$ do forward sweep\\
if $|L_{kk}| >  \rm zero$, then do:\\
$Q_{kk} \leftarrow \ominus_k(1/2)\times Q_{kk}/L_{kk}$,\\
$S_{kk} \leftarrow \pm_k 2 \times Q_{kk}F_{kk}$,\\
\textcolor{red}{$Q_{jk} \leftarrow [\ominus_k Q_{jk} - Q_{kk}L_{jk}]/L_{kk} $,}\\
$S_{jk} \leftarrow \pm_k Q_{kk}\times F_{jk}$,\\
$S_{kk} \leftarrow S_{kk}\pm_k(Q_{jk}\times F_{jk})$, for $j = k+1, ..., N$\\
$Q_{ij} \leftarrow Q_{ij} \pm_k (Q_{ik}\times L_{jk})\pm_k (L_{ik}Q_{jk})$,\\
$S_{ik} \leftarrow S_{ik}\pm_k(F_{ij}\times Q_{jk}), S_{jk}\leftarrow S_{jk}\pm_k (F_{ij}\times Q_{ik})$, for $j = k+1,...,N$ \& $i = j,...,N$\\
\\
\textcolor{red}{(b)  for $i \ge j, S_{ij} \leftarrow S_{ij} + \sum_{m \ge n} Q_{mn}\times \partial^2 F({\bf L})/\partial L_{mn} \partial L_{ij}$.}\\
(c) reverse sweep, ${\bf S} \leftarrow T({\bf S})$ (see Table 2). \\
\textcolor{red}{(d)  $\partial^2 F({\bf L})/\partial x_v \partial x_u = \sum_{i\ge j} S_{ij}\times \partial K_{ij}/\partial x_u + \sum_{i \ge j} F_{ij}\times \partial^2 K_{ij}/\partial x_v \partial x_u, u = 1, 2, ..., p$.}\\
\\\hline\\
Table 3. Pseudocode for Backward Differentiation Applied Twice to $F({\bf L})$.\\
\end{tabular}

\end{document}